\documentclass[reprint,amsmath,amssymb,aps]{revtex4-1}
\usepackage{graphicx}
\usepackage{dcolumn}
\usepackage{bm}
\usepackage{hyperref}

\newcommand{\dprod}{\displaystyle\prod}
\newcommand{\tprod}{\textstyle\prod}

\begin{document}

\title{Predictive Renormalization-Group Theory of Universality Classes in Nonlinear Systems}

\author{Ko Okumura}
 \altaffiliation{Department of Physics and Soft Matter Center, Ochanomizu University, 2-1-1, Ohtsuka, Bunkyo-ku, Tokyo 112-8610, Japan}

\date{\today}

\begin{abstract}

Universal scaling behavior appears across a wide range of nonlinear systems
despite substantial differences in their governing equations and physical
mechanisms. We develop a renormalization-group (RG) framework that
identifies two complementary RG mechanisms underlying such universality.
First, scale invariance generates RG fixed points corresponding to
asymptotic self-similar solutions. Second, repeated RG transformations
eliminate non-scale-invariant irrelevant structures, causing broad classes
of equations to flow toward the same fixed points and thereby form
universality classes. The framework applies to finite-time singularities,
long-time intermediate asymptotics, stochastic Edwards--Wilkinson growth,
nonlinear diffusion, density-dependent biological diffusion, and
fluid-interface dynamics. In each case, it reproduces known scaling behavior
and identifies the associated universality class through explicit
irrelevance criteria. A central feature of the framework is its predictive
character. Once a scale-invariant fixed point is identified, the theory
predicts whole classes of nonlinear equations sharing the same asymptotic
self-similar solution. While the diffusion class is partially supported by
existing mathematically rigorous results, most universality classes
identified here have not previously been established and therefore
constitute testable predictions. These results provide a unified RG
perspective on universality in nonlinear systems and show that universality
emerges from the same fundamental RG principles that underlie critical
phenomena.

\vspace{0.5em} \noindent\small Published in \textit{J. Chem. Phys.} \textbf{165}, 194116 (2026). Special Topic Honoring Prof.~Hajime Tanaka. DOI: \href{https://doi.org/10.1063/5.0356648}{10.1063/5.0356648}
\end{abstract}


\maketitle


\section{Introduction}

Self-similar behavior is one of the most remarkable forms of organization
observed in nature. It appears in phenomena as diverse as fluid breakup \cite%
{1994ScienceNagelDropFallingFaucet,Bertozzi1994} and coalescence \cite%
{YokotaPNAS2011, eggers2025coalescence}, the capillary leveling \cite%
{2014ScienceElieGlassyPolymer} or spreading \cite{Diez1994} of polymer
films, blast-wave propagation \cite{Barenblatt}, stochastic interface growth 
\cite{barabasi1995fractal}, and biological population dynamics \cite%
{murray2002mathematical1, murray2002mathematical2}. Despite profound
differences in physical mechanisms and governing equations, these systems
often develop characteristic scales that organize their dynamics into
universal asymptotic forms.

Over the past decades, powerful analytical approaches, particularly
similarity methods \cite{Barenblatt, barenblatt2003scaling} and the
dynamical-system description (DSD) \cite{eggers2015singularities} of
self-similar solutions, have provided profound insight into both finite-time
singularities and long-time \textit{intermediate asymptotics }\cite{Note1}.
Within the DSD, originating from the logarithmic-time formulation by Giga
and Kohn \cite{giga1985asymptotically} and extensively developed by others 
\cite%
{1993PRLEggersPinchoff,1994ScienceNagelDropFallingFaucet,brenner1994iterated,brenner1996pinching, Bertozzi1994, Diez1994}%
, self-similar solutions normally appear as attractors or fixed points (FPs)
of appropriately transformed dynamical systems.

While this stability analysis has successfully characterized a wide range of
phenomena, a renormalization-group (RG) interpretation provides a
complementary perspective on these asymptotic structures. In particular,
Goldenfeld \cite{Goldenfeld} explicitly demonstrated for nonlinear diffusion
that self-similar solutions can be identified with RG fixed points and
discussed the resulting connection with universality. Building on this
established RG picture, the present work addresses a more specific question:
how universality classes can be systematically and predictively constructed
from generic PDE structures by determining their scaling behavior and
identifying the structures that become irrelevant under repeated RG
transformations.

RG theory for critical phenomena \cite{Yeomans, Cardy, Goldenfeld}, for
which K. G. Wilson and P.-G. de Gennes were awarded the Nobel Prize in
Physics in 1982 and 1991, respectively, provides a fundamental framework for
understanding universality in systems with many degrees of freedom. The
Wilsonian RG theory and its concept of universality classes that emerge as
RG flows approach FPs have become central to modern physics, spanning hard 
\cite{altland2023condensed} and soft \cite{de1979scaling,
brochard2019essentials,doi2013soft,van2024soft} condensed matter and
non-equilibrium systems \cite{livi2017nonequilibrium, tailleur2022active}.

Motivated by the RG perspective provided for critical phenomena and by
repeated exposure to hierarchical universality in experiments (see Appendix %
\ref{A-hi}), the present author asked whether a similar RG principle
underlies the emergence of self-similar dynamics in nonlinear PDEs and
whether a counterpart of \textit{universality classes} exists for
self-similar solutions. He focused on two complementary RG approaches that
have previously been applied to deterministic PDEs \cite%
{Note2,martin1973statistical,Janssen1979, kardar1986dynamic}: the
field-theoretic RG theory \cite%
{goldenfeld1989intermediate,goldenfeld1990anomalous,Goldenfeld}, proposed by
physicists, Goldenfeld, Martin, and Oono, and the BKL-RG theory \cite%
{bricmont1994renormalization}, proposed by Bricmont, Kupiainen, and Lin. The
connection between the field-theoretic and iterative RG viewpoints had
already been demonstrated by Goldenfeld \cite{Goldenfeld} in the analysis of
nonlinear diffusion, where the iterative RG formulation reproduces the
asymptotic scaling obtained from the perturbative treatment. Following this
established connection, the present author unified the field-theoretic RG
and BKL-RG approaches with DSD stability analysis and applied the resulting
framework to several PDEs \cite%
{Okumura2025RG,okumura2026combined,Okumura2026oil}.

Building on these case-specific developments, the present work extends the
framework to a generic PDE and demonstrates for diverse nonlinear systems
that scale invariance generates self-similar solutions as RG FPs. The
framework further provides explicit predictions for broad classes of
governing equations that converge toward the same self-similar solution and
therefore belong to a common universality class. The present formulation
applies equally to finite-time singularities and long-time intermediate
asymptotics, stochastic growth processes, and biological diffusion systems,
demonstrating that various nonlinear systems are governed by the same RG
mechanism that underlies critical phenomena: the progressive elimination of
irrelevant structures toward common scale-invariant FPs.

Because this framework bridges concepts that have traditionally developed in
separate communities---including DSD, RG theory, and nonlinear PDE
analysis---an explanatory note is provided in Appendix \ref{A0} to
facilitate communication across disciplines.

\begin{figure}[tbp]
\includegraphics[width=\columnwidth]{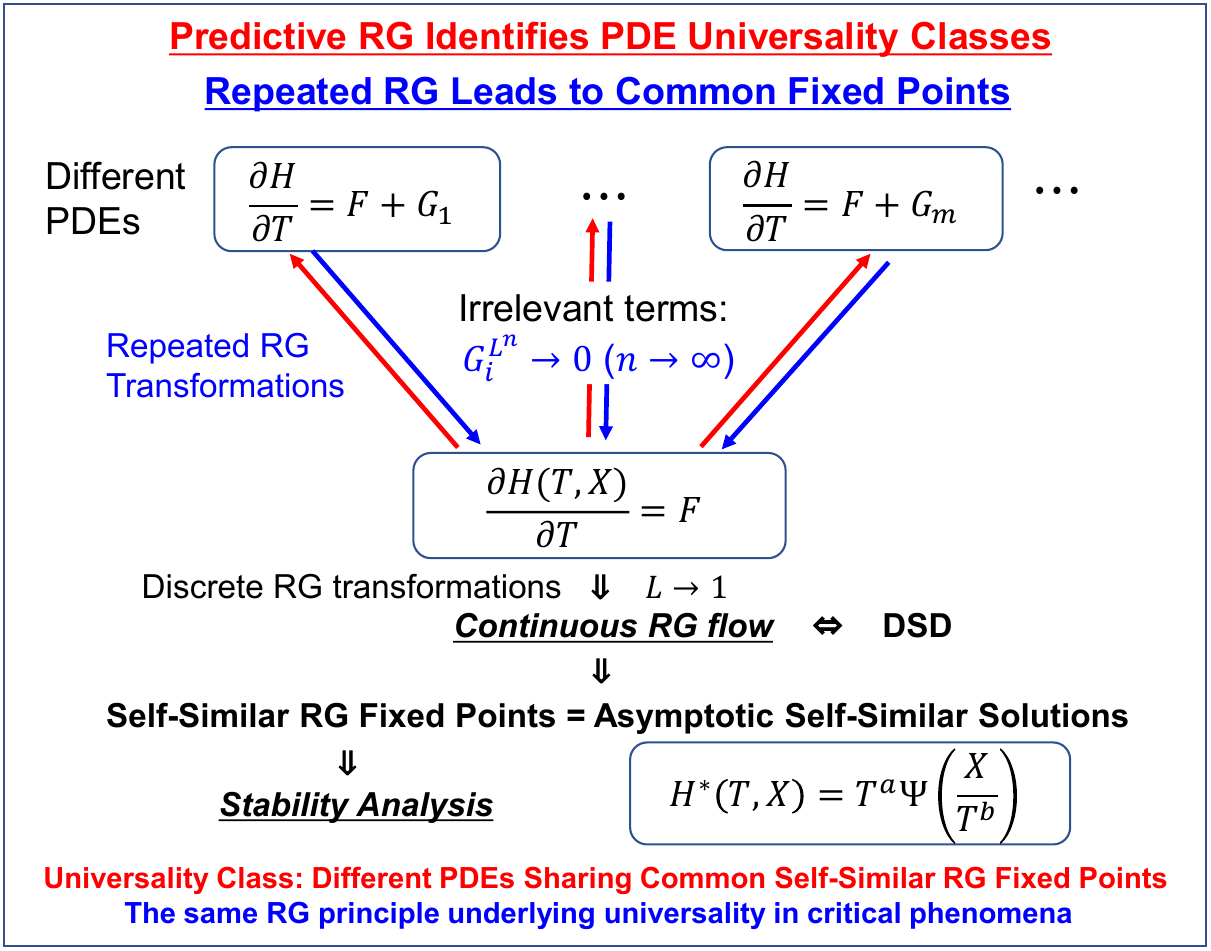}
\caption{Predictive RG framework for predicting universality classes through
repeated RG transformations. The red arrows illustrate the predictive use of
the framework: once a fixed-point equation is identified, the corresponding
universality classes formed by families of nonlinear PDEs can be predicted.}
\label{HI}
\end{figure}

The central concept of the present predictive RG framework is summarized
schematically in the Highlight Image (Fig \ref{HI}). The framework describes
how different nonlinear PDEs can flow toward common self-similar RG fixed
points through the progressive elimination of irrelevant structures and,
conversely, how the resulting universality classes can be predicted from the
identified fixed points. In this sense, universality classes themselves
become prediction targets of the RG framework. The Image also illustrates
the dynamical structure of the framework, connecting repeated discrete RG
transformations, continuous RG flow, the DSD, and the stability analysis of
the resulting fixed points, as discussed in detail in Sec. II B 5.

\section{Results}

\subsection{Self-similar solution and scaling ansatz}

The self-similar solution of a PDE can be expressed as%
\begin{equation}
H(T,X)=Y_{0}(T)\Gamma (X/X_{0}(T)),  \label{SS0}
\end{equation}%
with characteristic scales $Y_{0}(t)$ and $X_{0}(t)$. Based on Eq \ref{SS0},
profiles $Y=H(T,X)$ obtained at different times $T$ collapse onto a master
curve $\widetilde{Y}=\Gamma (\widetilde{X})$ when plotted on the rescaled
axes $\widetilde{X}=X/X_{0}(t)$ and $\widetilde{Y}=Y/Y_{0}(t)$. This
corresponds to the zeroth level of hierarchical universality introduced in
Appendix \ref{A-hi}. At this level, universality refers to the existence of
rescaling-induced data collapse itself, a property shared by many nonlinear
systems even when the governing equations and resulting self-similar
solutions are different.

The hierarchy of universality is organized according to the amount of
information shared by different observations. At the zeroth level, only data
collapse is required. At the first level, the asymptotic scaling function
and scaling exponents are additionally shared. At the second level, the same
asymptotic self-similar solution persists even when experimental conditions
are varied.

In many cases, the scales further satisfy scaling laws: $Y_{0}(t)\sim T^{a}$
and $X_{0}(t)\sim T^{b}$. In this case, the self-similar solution can be
written in the standard form 
\begin{equation}
H(T,X)=T^{a}\Gamma (X/T^{b}).  \label{SS}
\end{equation}%
The exponents $a$ and $b$ together with the scaling function $\Gamma (X)$
define a particular asymptotic self-similar solution, corresponding to the
first level of hierarchical universality. The second level arises when this
same asymptotic self-similar solution persists under variations of
experimental conditions.

The form in Eq \ref{SS} is mathematically equivalent to the scaling ansatz
in critical phenomena for (dimensionless) magnetization $M(t,h)$ as a
function of the magnetic field $h$ and temperature $t$ \cite%
{widom1965equation}:%
\begin{equation}
M(t,h)=t^{\beta }\Psi (h/t^{\Delta }),  \label{SA}
\end{equation}%
where $t\sim T_{c}-T$ and $h$ respectively correspond to temperature and
magnetic field with $T_{c}$ a critical temperature. The exponents $\beta $
and $\Delta $ are called the critical exponents and $\Psi $ is the scaling
function.

In critical phenomena, a universality class consists of substances and
models that share the same critical exponents $\beta $ and $\Delta $
together with the same asymptotic scaling function $\Psi (x)\sim x^{1/\delta
}$ in Eq \ref{SA}, under the conventional definitions of thermodynamic
quantities such as specific heat and susceptibility.

By mathematical analogy, we define a \textit{universality class} for PDEs as
a class of PDEs that share the same asymptotic self-similar solution in Eq %
\ref{SS}, characterized by the same scaling exponents $a$ and $b$ and the
same scaling function $\Gamma (X)$. This definition naturally extends the
second level of universality introduced in Appendix \ref{A-hi}, which
concerns the persistence of a common self-similar solution under variations
of experimental conditions within a given physical system. A universality
class generalizes the same concept to different governing equations.

\subsection{General theory}

\subsubsection{Generic governing equation}

We consider a set of PDEs in the following form for the function $\mathbf{H}%
=(H_{1},H_{2},\cdots ,H_{n})$ of the time variable $T$ and the spatial
variable $\mathbf{X}=(X_{1},X_{2},\cdots ,X_{m})$:

\begin{equation}
\partial _{T}\mathbf{H}(T;\mathbf{X})=\mathbf{F}(\mathbf{H},D_{1}\mathbf{H}%
,D_{2}\mathbf{H},\cdots ),  \label{c3}
\end{equation}%
where $D_{i}\mathbf{H}$ stands for the $n$-th spatial derivative, e.g., $%
D_{2}\mathbf{H}$ stands for the set $\{\partial _{X_{i}}\partial _{X_{j}}%
\mathbf{H}\}$. We assume that $\mathbf{F}$ is a \textit{regular analytic
function}, which here and hereafter means a convergent analytic function of $%
\mathbf{H}$ and a finite set of its derivatives of arbitrary but finite
order. A general term in $\mathbf{F}$ can be written as%
\begin{eqnarray}
&&H_{1}^{N_{1}}H_{2}^{N_{2}}\cdots (\partial
_{X_{1}}H_{1})^{N_{11}}(\partial _{X_{1}}H_{2})^{N_{12}}\cdots  \notag \\
&=&\dprod\limits_{i,j,k,l,n,\cdots }H_{i}^{N_{i}}(\partial
_{X_{j}}H_{k})^{N_{jk}}(\partial _{X_{l}}\partial
_{X_{m}}H_{n})^{N_{lmn}}\cdots  \label{t1}
\end{eqnarray}%
with the coefficient $\mathbf{f}_{\{N_{i}\},\{N_{jk}\},\{N_{lmn}\}\cdots }$
(see Appendix \ref{A2a} for full details). Equation \ref{t1} will be
exploited in Eq \ref{e1} below to determine the scaling exponent $M_{I}$
under the RG transformation.

A plethora of physical systems can be described in the generic form in Eq %
\ref{c3}.\ Representative examples span deterministic singular dynamics,
stochastic systems, biological population dynamics, chemical reactions, and
nonequilibrium pattern formation \cite{murray2002mathematical1,
cross1993pattern}. These include, for instance, the bubble PDE describing
the breakup of an inviscid fluid drop surrounded by a highly viscous fluid 
\cite{Okumura2025RG}: $\partial _{T}H(T,X)=1$, the linear diffusion
equation: $\partial _{T}H(T,X)=\partial _{X}^{2}H(T,X)/2$, the stochastic
Edwards-Wilkinson (EW) equation \cite%
{barabasi1995fractal,livi2017nonequilibrium}) describing stochastic
interface fluctuations: $\partial _{T}H(T,X)=\partial
_{X}^{2}H(T,X)/2+\Theta (T,X)$ where $\Theta (T,X)$ denotes Gaussian white
noise, the density-dependent (DD) diffusion model, which preserves the total
mass and serves as a representative example of biological nonlinear
diffusive systems: $\partial _{T}H(T,X)=D\partial _{X}(H^{m}\partial
_{X}H(T,X))$, and the oil-drop PDE describing the breakup in air of a
moderately viscous fluid, like olive oil \cite{1993PRLEggersPinchoff}:%
\begin{equation}
\left\{ 
\begin{array}{c}
-\frac{\partial H}{\partial T}+V\frac{\partial H}{\partial X}=-\frac{H}{2}%
\frac{\partial V}{\partial X} \\ 
-\frac{\partial V}{\partial T}+V\frac{\partial V}{\partial X}=-\frac{%
\partial }{\partial X}\left( \frac{1}{H}\right) +3\frac{1}{H^{2}}\frac{%
\partial }{\partial X}\left( \frac{\partial V}{\partial X}H^{2}\right)%
\end{array}%
\right.  \label{G3D}
\end{equation}%
where $H$ stands for the function describing the profile of the interface
and $V$ stands for the radial component of the velocity of the viscous
fluid. These examples illustrate the generality of the present RG framework
across fundamentally different classes of linear and nonlinear systems.

We are interested in \textit{singular dynamics} or \textit{intermediate
asymptotics} with the boundary condition at $T=1$:%
\begin{equation}
\mathbf{H}(1,\mathbf{X})=\mathbf{h}(\mathbf{X})  \label{eq-2}
\end{equation}

To investigate how additional structures affect the asymptotic dynamics of
Eq \ref{c3}, we further consider a broader class of PDEs of the form:%
\begin{equation}
\partial _{T}\mathbf{H}(T;\mathbf{X})=\mathbf{F}(\mathbf{H},D_{1}\mathbf{H}%
,D_{2}\mathbf{H},\cdots )+\mathbf{G}(\mathbf{H},D_{1}\mathbf{H},D_{2}\mathbf{%
H},\cdots ),  \label{e6}
\end{equation}%
where $\mathbf{F}$ will be taken to represent the scale-invariant part of
the governing equation, in the sense defined below, whereas $\mathbf{G}$
collects additional terms whose RG relevance or irrelevance will be
determined from their scaling behavior. As shown below, irrelevant terms in $%
\mathbf{G}$ are eliminated under repeated RG transformations, providing the
basis for the construction of universality classes. As for $\mathbf{F}$, $%
\mathbf{G}(\mathbf{H},D_{1}\mathbf{H},D_{2}\mathbf{H},\cdots )$ is assumed
to be a regular analytic function of $\mathbf{H}$ and its spatial
derivatives, defined through the coefficients $\mathbf{g}_{\{N_{i}\},%
\{N_{jk}\},\{N_{lmn}\}\cdots }$.

\subsubsection{Scale transformation}

Following the original article \cite{bricmont1994renormalization}, we
introduce the scale transformation: 
\begin{eqnarray}
T^{\prime } &=&T/L^{B},\text{ }\mathbf{X}^{\prime }=\mathbf{X}/L  \label{Ar3}
\\
H_{i}^{\prime }(T^{\prime };\mathbf{X}^{\prime }) &=&L^{A_{i}}H_{i}(T;%
\mathbf{X})\text{ }\equiv H_{i}^{L}(T^{\prime },\mathbf{X}^{\prime })  \notag
\label{Ar3b} \\
&\Leftrightarrow &\text{ }H_{i}^{L}(T;\mathbf{X})=L^{A_{i}}H_{i}(L^{B}T;L%
\mathbf{X}).  \label{Ar3c}
\end{eqnarray}%
\newline
where $L$ and $B$ are assumed to be positive. Note here that the exponents $%
A_{i}$ and $-B$ are counterparts of \textit{the scaling dimension}
(respectively for $H_i(T;{\mathbf{X}})$ and $T$) in the RG for critical
phenomena \cite{Note3}.

For self-similar solutions, two opposite asymptotic regimes are important:
One corresponds to finite-time singular dynamics in the limit $T\rightarrow
0 $, which may be viewed as a short-time \textit{intermediate asymptotics}
in the terminology of Barenblatt. The other corresponds to long-time
intermediate asymptotics in the limit $T\rightarrow \infty $. Accordingly,
we consider two cases: (I) $0<L<1$ and (II) $L>1$. In Case I, the scale
transformation in Eq \ref{Ar3} corresponds to magnification: the new
variables $T^{\prime }$ and $\mathbf{X}^{\prime }$ are larger than $T$ and $%
\mathbf{X}$. On the contrary, in Case II, the transformation corresponds to
de-magnification. Note that the BKL-RG was proposed for Case II and extended
for Case I in the context of gravitational collapse of stars \cite%
{koike1995critical}.

We require Eq \ref{c3} to be invariant under the scale transformation so
that Eqs \ref{c3} and \ref{e6} respectively transform into the following
forms: 
\begin{eqnarray}
\partial _{T}\mathbf{H}_{L}(T;\mathbf{X}) &=&\mathbf{F}(\mathbf{H}_{L},D_{1}%
\mathbf{H}_{L},D_{2}\mathbf{H}_{L},\cdots )  \label{e7a} \\
\partial _{T}\mathbf{H}_{L}(T;\mathbf{X}) &=&\mathbf{F}(\mathbf{H}_{L},D_{1}%
\mathbf{H}_{L},D_{2}\mathbf{H}_{L},\cdots )  \notag \\
&&+\mathbf{G}_{L}(\mathbf{H}_{L},D_{1}\mathbf{H}_{L},D_{2}\mathbf{H}%
_{L},\cdots ).  \label{e7b}
\end{eqnarray}

For example, the term of the form in Eq \ref{t1} in $F_{I}$ (the $I$-th
component of $\mathbf{F}$) or $G_{I}$ transforms as $L^{M_{I}}\tprod%
\limits_{i,j,k,l,n,\cdots }\left[ H_{i}^{L}\right] ^{N_{i}}(\partial
_{X_{j}}H_{k}^{L})^{N_{jk}}(\partial _{X_{l}}\partial
_{X_{m}}H_{n}^{L})^{N_{lmn}}\cdots $ with the \textit{scaling factor} $%
L^{M_{I}}$ characterized by the \textit{scaling exponent }$M_{I}$:%
\begin{eqnarray}
M_{I} &=&A_{I}+B-[\sum_{i}N_{i}A_{i}+\sum_{j,k}N_{jk}(A_{k}+1)  \notag \\
&&+\sum_{l,m,n}N_{lmn}(A_{n}+2)+\cdots ],  \label{e1}
\end{eqnarray}%
as explained in Appendix \ref{A-D1}. This expression plays a central role in
the present framework, as it provides a systematic criterion for identifying
scale invariance and determining the irrelevance of the extra term $\mathbf{G%
}$, as will be illustrated below.

From Eq \ref{e1}, we can understand (A) the scale invariance of Eq \ref{c3}
means that the coefficient $f_{\{N_{i}\},\{N_{jk}\},\{N_{lmn}\}\cdots }^{I}$
is nonzero only when the scaling exponent $M_{I}$ for the corresponding term
is zero ($M_{I}$ for the term with nonzero coefficient should be zero since
the transform of the term should be independent of the scaling factor $L$),
and (B) the term in $\mathbf{G}_{L}(\mathbf{H},D_{1}\mathbf{H},D_{2}\mathbf{H%
},\cdots )$ with positive (negative) $M_{I}$ vanishes as $L\rightarrow 0$
for Case I ($L\rightarrow \infty $ for Case II). This means that, with
defining the \textit{scaling exponent} $N_{G}$ as the smallest (largest) of $%
M_{I}$ for the term with the nonzero coefficient $g_{\{N_{i}\},\{N_{jk}\},%
\{N_{lmn}\}\cdots }^{I}$ for Case I (Case II), the regular analytic function 
$\mathbf{G}_{L}(\mathbf{H},D_{1}\mathbf{H},D_{2}\mathbf{H},\cdots )$ is 
\textit{irrelevant} if $N_{G}$ is positive (negative) for Case I (Case II)
in the following sense: the term disappears as $L\rightarrow 0$ for Case I ($%
L\rightarrow \infty $ for Case II).

\subsubsection{RG transformation}

Now the RG transformation is defined as%
\begin{equation}
\emph{R}_{L,\mathbf{G}}h_{i}(\mathbf{X})\equiv L^{A_{i}}H(L^{B};L\mathbf{X}%
)=H_{i}^{L}(1;\mathbf{X})\text{ },
\end{equation}%
where $h_{i}(\mathbf{X})$ is the $i$-th component of $\mathbf{h}(\mathbf{X})$
in Eq \ref{eq-2}. The second equality is based on Eq \ref{Ar3c}. This RG
transformation can be regarded as two step, as in Wilson's RG theory. In the
present case, the first step is the time evolution based on Eq \ref{e6},
from $T=1$ to $T=L^{B}$ ($L^{B}$ is smaller and larger than $1$ respectively
for Case I and II), because of which $\emph{R}_{L,\mathbf{G}}$ depends on $%
\mathbf{G}$: $\mathbf{h}(\mathbf{X})=\mathbf{H}(1;\mathbf{X})\overset{%
\mathbf{G}}{\rightarrow }\mathbf{H}(L^{B};\mathbf{X})$. The second step is
the rescaling of $\mathbf{X}$ and $\mathbf{H}$: $H_{i}(L^{B};\mathbf{X}%
)\rightarrow L^{A_{i}}H_{i}(L^{B};L\mathbf{X})$.

If we repeat the scale transformation, the time and space variables change
as $T$ $\rightarrow $ $T/L^{B}$ $\rightarrow $ $T/L^{2B}$ $\rightarrow $ $%
\cdots $ $\rightarrow T/L^{kB}$ and $\mathbf{X}\rightarrow $ $\mathbf{X}%
/L\rightarrow \mathbf{X}/L^{2}\rightarrow \cdots $ $\rightarrow $ $\mathbf{X}%
/L^{k}$: the space-time is progressively magnified (de-magnified) for Case I
(Case II), while the extra term $\mathbf{G}$ changes as $\mathbf{G}$ $%
\rightarrow $ $\mathbf{G}_{L}\rightarrow $ $\mathbf{G}_{L^{2}}$ $\rightarrow 
$ $\cdots $ $\rightarrow $ $\mathbf{G}_{L^{k}}$. Thus, the initial function
changes as $\mathbf{H}(1;\mathbf{X})\overset{\mathbf{G}}{\rightarrow }%
\mathbf{H}_{L}(1;\mathbf{X})\overset{\mathbf{G}_{L}}{\rightarrow }\mathbf{H}%
_{L^{2}}(1;\mathbf{X})\overset{\mathbf{G}_{L^{2}}}{\rightarrow }\cdots 
\overset{\mathbf{G}_{L^{k-1}}}{\rightarrow }\mathbf{H}_{L^{k}}(1;\mathbf{X})$%
. After repeating the RG transformation $m$ times, the function $\mathbf{h}(%
\mathbf{X})=\mathbf{H}(1;\mathbf{X})$ representing the initial condition is
transformed into $\mathbf{H}_{L^{m}}(1;\mathbf{X})$. The ($m+1$)-th RG
operation applied to this function is given by $\emph{R}_{L,\mathbf{G}%
_{L^{m}}}\mathbf{H}_{L^{m}}(1;\mathbf{X})$ $=L_{L^{m}}^{A}\mathbf{H}%
(L^{B},LX)$ [$=\mathbf{H}_{L^{m+1}}(1;\mathbf{X})$] (An important point
concerning the iteration of this relation is given in Note \cite{Note4}). In
other words, the time evolution from $T=1$ to $T=L^{B}$ in this case is
governed by $\mathbf{G}_{L^{m}}$. Thus, the repetition of the RG
transformation is defined as%
\begin{equation}
\emph{R}_{L^{n}}\mathbf{h}(\mathbf{X})=\emph{R}_{L,\mathbf{G}%
_{L^{n-1}}}\circ \cdots \emph{R}_{L,\mathbf{G}_{L}}\circ \emph{R}_{L,\mathbf{%
G}}\mathbf{h}(\mathbf{X}).  \label{eq-4b}
\end{equation}

Suppose that after $m$ iterations $\mathbf{H}_{L^{m}}(1;\mathbf{X})$ reaches
a FP; i.e., $\mathbf{H}_{L^{m+1}}(1;\mathbf{X})=\mathbf{H}_{L^{m}}(1;\mathbf{%
X})\equiv \mathbf{H}^{\ast }(1;\mathbf{X})$. This implies that the equation
governing the ($m+1$)-th RG transformation, i.e., 'Eq \ref{e7b} with $L$
replaced by $L^{m}$,' has also reached a FP. In other words, for a FP to
exist, the governing equation (\ref{e6}) must become scale-invariant. This
requires Eq \ref{e7b}\ to be a form independent of $L$, which demands $%
\mathbf{G}_{L^{m+1}}=\mathbf{G}_{L^{m}}\equiv \mathbf{G}^{\ast }$ and the
scale invariance of Eq \ref{c3}. In such a case, we can expect that $\emph{R}%
_{L^{n}}\mathbf{h}(\mathbf{X})$ and 'Eq \ref{e7b} with $L$ replaced by $%
L^{n} $' flow into their FPs: $\mathbf{h}^{\ast }(\mathbf{X})$ and 'Eq \ref%
{e6} with $\mathbf{G}=\mathbf{G}^{\ast }$.' At this stage, the fixed-point
contribution $\mathbf{G}^{\ast }$ need not generally vanish. In such a case,
Eq \ref{c3} must be scale invariant and the FP of the RG transformation is
defined by the following equation:%
\begin{equation}
\emph{R}_{L,\mathbf{G}^{\ast }}\mathbf{h}^{\ast }(\mathbf{X})=\mathbf{h}%
^{\ast }(\mathbf{X})\text{ $\Leftrightarrow $ }L^{A_{i}}H_{i}^{\ast }(L^{B};L%
\mathbf{X})=h_{i}^{\ast }(\mathbf{X}).  \label{eq-5}
\end{equation}%
If such a point exists, setting $T=L^{B}$ in this equation results in $%
T^{A_{i}/B}H_{i}(T;T^{1/B}\mathbf{X})=\mathbf{h}^{\ast }(\mathbf{X})$, from
which we obtain $H_{i}^{\ast }(T;\mathbf{X})=T^{-A_{i}/B}h_{i}^{\ast }(%
\mathbf{X}/T^{1/B})$. In other words, \textit{self-similar solutions emerge
as FPs if }$\mathbf{h}^{\ast }(\mathbf{X})$ \textit{exists}: 
\begin{equation}
H_{i}^{\ast }(T;\mathbf{X})=T^{-A_{i}/B}h_{i}^{\ast }(\bm{\xi })\text{ with }%
\bm{\xi }=\mathbf{X}/T^{1/B}  \label{e20}
\end{equation}%
where $\mathbf{h}^{\ast }(\bm{\xi })$ corresponds to the master curve for
the self-similar solution.

\subsubsection{Universality class\label{A2-7}}

We now return to the role of $\mathbf{G}$ introduced in Eq \ref{e6} and
specialize the general fixed-point argument above to the case in which $%
\mathbf{G}$ is irrelevant (i.e., ${\mathbf{G}}_{L^{n}}\sim
L^{nN_{G}}\rightarrow 0$ as $n\rightarrow \infty $) and thus $\mathbf{G}%
^{\ast }=0$. In this case, repeated RG transformations drive $\mathbf{G}_{L}$%
{} toward its fixed-point value, $\mathbf{G}^{\ast }=0$. The exponent $N_{G}$%
{} was introduced precisely to characterize this property. Consequently, if $%
N_{G}${} is positive (negative) for Case I (Case II), there is no difference
in the resulting asymptotic self-similar solution obtained by searching for
an RG fixed point, regardless of whether $\mathbf{G}$ is initially present
or absent. Thus, an additional term is called irrelevant when it satisfies
this positivity (negativity) condition.

In other words, a broad class of PDEs, including complex and nonlinear
equations that can be expressed in the form of Eq \ref{e6} with an
irrelevant $\mathbf{G}$, flows toward the same RG fixed point as the simpler
Eq \ref{c3}. These equations therefore share the same asymptotic
self-similar solution and constitute the universality class of Eq \ref{c3}.

\subsubsection{RG flow equation and stability analysis\label{A2-3 copy(1)}}

From the conclusion of the previous section, to find $\mathbf{h}^{\ast }(%
\mathbf{X})$ to explicitly obtain FPs (which correspond to self-similar
solutions) shared by the universality class, it is sufficient to set $%
\mathbf{G}=\mathbf{0}$; we shall assume this hereafter. Furthermore, we
introduce the logarithmic time $\tau =-B\log L$ for Case I. When the RG
scale factor $L$ is associated with the time variable through $T=L^{B}$,
this gives $\tau =-\log T$; hence, $\tau \rightarrow \infty $ corresponds to 
$T\rightarrow 0$. For Case II, we introduce $\tau =B\log L$ instead, which
guarantees that $\tau $ is positive and goes to infinity as $T\rightarrow
\infty $, as in Case I.

Then, the following equation can be derived by using the relation $\mp
B/d\tau =L/dL$ and Eq \ref{Ar3c}: 
\begin{eqnarray}
\mp B\frac{dH_{i}^{L}(T;\mathbf{X})}{d\tau } &=&A_{i}H_{i}^{L}(T;\mathbf{X}%
)+\sum_{i}X_{i}\frac{\partial H_{i}^{L}(T;\mathbf{X})}{\partial X_{i}} 
\notag \\
&&+BTF_{i}(\mathbf{H}_{L},D_{1}\mathbf{H}_{L},D_{2}\mathbf{H}_{L},\cdots ),
\label{e10}
\end{eqnarray}%
where the minus and plus signs on the left-hand side correspond to Case I
and II, respectively. By setting $T=1$ in this equation and introducing $%
\widehat{\mathbf{h}}(\tau ;\mathbf{X})$ by the relation $\mathbf{H}_{L}(1;%
\mathbf{X})=\emph{R}_{L}\mathbf{f}(\mathbf{X})\equiv \widehat{\mathbf{h}}%
(\tau ;\mathbf{X})$, we obtain 
\begin{eqnarray}
\mp B\frac{d\widehat{h_{i}}(\tau ;\mathbf{X})}{d\tau } &=&A_{i}\widehat{h_{i}%
}(\tau ;\mathbf{X})+BF_{i}(\widehat{\mathbf{h}},D_{1}\widehat{\mathbf{h}}%
,D_{2}\widehat{\mathbf{h}},\cdots )  \notag \\
&&+\sum_{i}X_{i}\frac{\partial \widehat{h_{i}}(\tau ;\mathbf{X})}{\partial
X_{i}}.  \label{e11}
\end{eqnarray}%
This is the RG 'flow'\ equation. Since changes in $\tau $ correspond to the
changes (in $L$) occurring as the RG transformation is repeatedly applied
(see Note \cite{Note5} for further explanations), this equation governs the
'flow'\ associated with the RG transformation. Therefore, we can obtain FPs, 
$\widehat{\mathbf{h}}(\tau ;\mathbf{X})=\mathbf{h}^{\ast }(\mathbf{X})$,
from Eq \ref{e11} by setting $\frac{d\widehat{h_{i}}(\tau ;\mathbf{X})}{%
d\tau }=0$ and perform stability analysis around the FP, $\mathbf{h}^{\ast }(%
\mathbf{X})$: we substitute 
\begin{equation}
\widehat{\mathbf{h}}(\tau ;\mathbf{X})=\mathbf{h}^{\ast }(\mathbf{X})+%
\bm{\delta}(\mathbf{X})e^{\omega \tau }
\end{equation}%
into Eq \ref{e11}, linearize the equation in terms of $\bm{\delta}(\mathbf{X}%
)$, and examine the sign of $\omega $. Since $\tau $ goes to positive
infinity for both Cases of I and II, a negative $\omega $ corresponds to a 
\textit{stable} (decaying) mode, whereas a positive $\omega $ indicates an 
\textit{unstable} (growing) mode. The case $\omega =0$ corresponds to a 
\textit{marginal} mode, which may be absorbed into the FP solution depending
on the structure of the problem as seen in Appendix \ref{A3}.

Such stability analyses have been performed frequently for singular dynamics
and called \textit{dynamical-system description} (DSD). In the conventional
DSD, the counterpart of Eq \ref{e11} was ingeniously derived for $n=m=1$ by
simply assuming the existence of self-similar solutions and further assuming
a generalized self-similar form, $H(T,X)=T^{\alpha }\widehat{f}(\tau ,\xi )$
with $\xi =X/T^{\beta }$, without discussing scale transformations, their
invariance, or defining RG transformations. Consequently, while the DSD
framework robustly analyzes the mathematical characteristics of these
solutions, a clear physical picture regarding the origin of the existence of
self-similar solutions---as elucidated in the above RG theory---remains
outside the scope of the DSD. In contrast, the present RG theory unifies
DSD, giving a physical meaning to the flow equation and FPs. They are RG
flow equations and RG FPs. In this manner, the present unified RG theory
integrates the DSD stability analysis.

\subsection{Application of the unified framework}

The four exemplary PDEs discussed below illustrate complementary regimes
encompassed by the unified framework, with additional two more PDEs
discussed in Appendix \ref{A2b}. Together, all the six exemplary PDEs span
finite-time singularities, long-time intermediate asymptotics, anomalous
scaling, and self-similarity of the first and second kind, while also
demonstrating the applicability of the framework to deterministic,
stochastic, and biological systems.

\subsubsection{Example 1: Bubble PDE [$\partial _{T}H(T,X)=1$]}

This seemingly simple yet non-trivial PDE, whose physical validity is
established by definitive experiments \cite%
{2003ScienceNagelMemoryDropBreakup,pahlavan2019restoring}, provides an
illustration of how a stable self-similar solution is obtained as an RG FP,
demonstrating that a wide class of nonlinear equations flows toward this
identical scaling solution to form a single universality class. In this case
($L<1$ and $n=m=1$), as explained in Appendix \ref{A2b}, $M_{I}$ ($A_{I}$)
has only one component $M$ ($A$), which is given as $M=A+B$ from Eq \ref{e1}%
; We obtain $A+B=0$ from the scale-invariance condition for Eq \ref{c3} and
the RG FPs (the self-similar solution) from Eq \ref{e20}: $H^{\ast
}(T,X)=T^{-A/B}h^{\ast }(X/T^{1/B})$. From Eq \ref{e11}, the RG flow
equation becomes%
\begin{equation}
-B\frac{d\widehat{h}(\tau ,X)}{d\tau }=-B\widehat{h}(\tau ,X)+B+X\frac{%
\partial \widehat{h}(\tau ,X)}{\partial X}.  \label{RGF}
\end{equation}%
We set the left-hand side to zero to have a 'stationary' FP solution: $%
\widehat{h}(\tau ,X)=h^{\ast }(X)$ with $h^{\ast }(X)=1+CX^{B}$. We require
a regularity condition and select a stable solution as experimentally
observed one in the stability analysis based on Eq \ref{RGF}, to conclude $%
B=2=-A$ (see Refs. \cite{Okumura2025RG,eggers2015singularities} for the
details). Then, from Eq \ref{e20}, the experimentally relevant, stable RG FP
solution is given by $H^{\ast }(T,X)=Th^{\ast
}(X/T^{1/2})=T[1+C(X/T^{1/2})^{2}]$ \cite{Note6}.

The present framework predicts that any PDE containing only irrelevant
additional terms with $M>0$ ($M<0$) for Case I (Case II) should approach the
same asymptotic fixed point. As an example, as shown in Appendix \ref{A2b},
the \textit{bubble breakup universality class} is revealed to consist of the
PDEs of the form $\partial _{T}H(T,X)=1+G$, where the regular analytic term $%
G$ is given by a linear combination of terms that can be expressed as $%
H^{n_{0}}$ $(\partial _{X}H)^{n_{1}}$ $(\partial _{X}^{2}H)^{n_{2}}$ $%
(\partial _{X}^{3}H)^{n_{3}}\cdots $ with a \textit{positive }$%
M=2n_{0}+n_{1}-(n_{3}+2n_{4}+3n_{5}\cdots )$ (Our reasoning for the
irrelevance of the extra term $G$ remains valid even when the exponents $%
n_{0},n_{1},\ldots $ are non-integer). As an example, a simple \textit{%
nonlinear} PDE where $G$ is a linear combination of $H^{n}$ with $n>0$
shares the same self-similar solution with $\partial _{T}H(T,X)=1$,
illustrating the broad scope of the universality class (as in conventional
RG theory, this statement concerns solutions belonging to the basin of
attraction of the fixed point).

\subsubsection{Example 2: Linear Diffusion [$\partial _{T}H(T,X)=\partial
_{X}^{2}H(T,X)/2$]}

In this case ($L>1$ and $n=m=1$), as explained in Appendix \ref{A2b}, $%
M=A+B-(A+2)$, which should be zero from the scale-invariance condition,
giving $B=2$ and the RG FPs: $H^{\ast }(T,X)=T^{-A/2}h^{\ast }(X/T^{1/2})$.
From Eq \ref{e11}, the RG flow equation becomes%
\begin{equation}
2\frac{d\widehat{h}(\tau ,X)}{d\tau }=A\widehat{h}(\tau ,X)+\frac{\partial
^{2}\widehat{h}(\tau ,X)}{\partial X^{2}}+X\frac{\partial \widehat{h}(\tau
,X)}{\partial X}  \label{FL1}
\end{equation}%
By setting the left-hand side to zero, we have a stationary FP solution: $%
H_{L}(T,X)=h^{\ast }(X)$ at $A=1$ with $h^{\ast }(X)=Ce^{-X^{2}/2}$. (The
conclusion $A=1$ can also be obtained by requiring the scale invariance of
the mass conserving equation $\int_{-\infty }^{\infty }dXH(T,X)=const.$)
This gives the self-similar solution (short-time intermediate asymptotics): $%
H^{\ast }(T,X)=CT^{-1/2}e^{-X^{2}/(2T)}$.

To make contact with the DSD, the present author performed a stability
analysis in this case \cite{okumura2026combined} by introducing a deviation: 
$\widehat{h}(\xi ,\tau )=f^{\ast }(\xi )+\delta f(\xi ,\tau )$ with $\delta
f(\xi ,\tau )=\delta (\xi )e^{\omega \tau }$. Substituting this into Eq~\ref%
{FL1}, we obtain an \textit{eigenvalue problem} for $\delta (\xi )$, whose
spectrum confirms that all eigenvalues $\omega $ are practically negative,
establishing the stability of the FP solution (see Appendix \ref{A3} for
further details).

The framework predicts that the \textit{diffusion universality class} can be
shown to consist of PDEs of the form $\partial _{T}H(T,X)=\partial
_{X}^{2}H(T,X)/2+G$ where the regular analytic term $G$ is given by a linear
combination of terms that can be expressed as $H^{n_{0}}(\partial
_{X}H)^{n_{1}}(\partial _{X}^{2}H)^{n_{2}}(\partial _{X}^{3}H)^{n_{3}}\cdots 
$ with a \textit{negative} $M=3-(n_{0}+2n_{1}+3n_{2}+\cdots )$. As an
example, a simple nonlinear PDE where $G$ is a linear combination of $H^{n}$
with $n>3$ shares the same self-similar solution with $\partial
_{T}H(T,X)=\partial _{X}^{2}H(T,X)/2$, illustrating the broad scope of the
universality class. This prediction, in the case of integer exponents, has
been rigorously established by Bricmont, Kupiainen, and Lin \cite%
{bricmont1994renormalization}, whereas the non-integer case still awaits
further numerical and mathematical verification.

It is worth noting that this framework can be extended to \textit{stochastic
systems}, provided that the constituent terms transform independently under
the RG evolution. A prime example is the noisy EW equation, where the noise
term itself remains scale-invariant, unlike the Kardar-Parisi-Zhang (KPZ)
equation, where the nonlinear term induces anomalous dimensions in both the
diffusion and noise terms \cite{barabasi1995fractal}. Consequently, the
present unified RG method can be applied directly to stochastic systems,
successfully reproducing the scaling exponents of the EW equation and
thereby identifying the explicit condition satisfied by PDEs forming the
corresponding universality class. The applicability of the framework to both
the EW equation and the DD diffusion model is discussed further in Appendix %
\ref{A2b}. Importantly, the DD diffusion model provides a representative
biological example to which the same RG framework can be applied. Notably,
these two additional cases further illustrate the predictive power of the
framework by yielding additional universality-class predictions.

\subsubsection{Example 3: Barenblatt PDE (Nonlinear Diffusion)}

Barenblatt discussed the long time asymptotic solution to $\partial
_{T}H(T,X)=D\partial _{X}^{2}H(T,X)$ with the initial condition $f(X)=\frac{%
m_{0}}{\sqrt{2\pi l^{2}}}\exp \left( -\frac{X^{2}}{2l^{2}}\right) $,
characterized by $m_{0}$ and $l$, at $T=0$, where the diffusion constant $D$
exhibits a step-like jump depending on the sign of the temporal rate of
change $\partial _{T}H$: there exists a domain $-X_{0}(T)<X<X_{0}(T)$ inside
of which $\partial _{T}H$ is negative and outside of which $\partial _{T}H$
is positive such that

\begin{equation}
D=\left\{ 
\begin{array}{cc}
\frac{1}{2}+\frac{\varepsilon }{2} & \text{for the inside:}%
-X_{0}(T)<X<X_{0}(T) \\ 
\frac{1}{2} & \text{ for the outside: }X<-X_{0}(T)\text{, }X_{0}(T)<X%
\end{array}%
\right.
\end{equation}

In this case ($L>1$ and $n=1$), we again have $B=2$ and $H^{\ast
}(T,X)=T^{-A/2}h^{\ast }(X/T^{1/2})$. The RG flow equation is slightly
modified as the coefficient of second term in the right-hand side of Eq~\ref%
{FL1} is replaced by 2$D$.

Goldenfeld, Martin, and Oono revisited the Barenblatt problem in pioneering
work on field-theoretic RG for deterministic PDEs \cite%
{goldenfeld1989intermediate,goldenfeld1990anomalous,Goldenfeld}, which led
to subsequent developments involving mathematicians \cite%
{chen1994renormalization,chen1995numerical,paquette1994structural,aronson1994calculation,ziane2000certain,deville2008analysis}%
. They performed an expansion in the small parameter $\varepsilon $ (i.e., $%
H=H_{0}+\varepsilon H_{1}+\cdots $), which yields a first-order term $H_{1}$
containing a logarithmic divergence of the form $\log (T/l^{2})$, analogous
to those in quantum field theory. This term diverges as the regularization
parameter $l$ (the initial width) tends to zero, or as the time $T$ tends to
infinity, rendering the naive series expansion invalid for long-time
asymptotics. They treated this divergence by exploiting renormalization
techniques developed in quantum field theory \cite{ryder1996quantum}, based
on the fundamental idea underlying the 1965 Nobel-prize work of Tomonaga,
Schwinger, and Feynman, forming a cornerstone of modern particle physics:
They introduced a renormalization constant $Z$ to absorb the singular part
and thereby effectively re-summing the divergent terms into a finite form, $%
T^{-(1/2+\alpha )}$ $\sim T^{-1/2}$ $(1-\alpha \log (T)+\cdots )$, where $%
\alpha $ is called the anomalous dimension. This remarkable achievement
allowed for the analytical determination of $\alpha $, although
perturbative, that had not been available (see Ref \cite%
{aronson1994calculation} for further development for the analytical results).

In the course, they obtain the following result: $H^{\ast }(X,T)=\frac{%
m_{0}e^{-\frac{X^{2}}{2T}}}{\sqrt{2\pi T}}\left[ 1-\alpha \log \frac{T}{l^{2}%
}+\cdots \right] $, which, to first order in $\alpha ,$ can be written as

\begin{eqnarray}
H^{\ast }(X,T) &=&\frac{T^{-\frac{1}{2}-\alpha }m_{0}l^{2\alpha }e^{-\frac{%
X^{2}}{2T}}}{\sqrt{2\pi }}  \label{H0} \\
&=&T^{-\frac{1}{2}-\alpha }h^{\ast }(X/T^{1/2})  \label{H}
\end{eqnarray}%
with $\alpha =\frac{\varepsilon }{\sqrt{2\pi e}}+O(\varepsilon ^{2})$ and $%
h^{\ast }(X)=Ce^{-X^{2}/2}$ with $C=\frac{m_{0}l^{2\alpha }}{\sqrt{2\pi }}$.

An important physical aspect of this result, emphasized by Barenblatt and
Goldenfeld \cite{Barenblatt, Goldenfeld}, is that the anomalous dimension is
accompanied by retention of information about the initial length scale.
Equation (\ref{H0}) makes explicit the retention of information about the
initial length scale $l$. For ordinary diffusion ($\alpha =0$), the
conserved mass $m_{0}${} remains in the long-time asymptotic solution,
whereas the initial width $l$ disappears. In the Barenblatt problem ($\alpha
\neq 0$), by contrast, information about $l$ survives through the
combination $m_{0}l^{2\alpha }$. Thus, the anomalous dimension is directly
associated with the incomplete loss of initial-scale information \cite%
{Goldenfeld}. In this specific sense, the Barenblatt solution retains memory
of the initial scale. A broader RG treatment of memory retention is
discussed in Ref \cite{Okumura2026memory}.

The retention of initial-condition information in asymptotic scaling is not
restricted to the Barenblatt model. An instructive experimental example was
provided by Diez, Gratton, and Gratton \cite{diez1992self, verga2003elements}%
, who studied a converging viscous gravity current approaching finite-time
focusing. In their self-similar solution, the nontrivial similarity exponent
is determined by the similarity problem, whereas a dimensionless coefficient
characterizing the asymptotic solution remains dependent on the initial
condition. Their experiments confirmed the predicted self-similar behavior
and similarity exponent. Thus, this system provides a physical example in
which detailed information about the initial state is largely eliminated
while specific information remains encoded in the asymptotic scaling
solution.

The temporal scaling in (\ref{H}) gives $A=1+2\alpha $. Moreover,
consistency between the limits $T\rightarrow \infty $ at fixed $l$ and $%
l\rightarrow 0$ at fixed $T$ requires the combination $m_{0}l^{2\alpha }$ to
remain finite in the small $l$ limit, which is naturally interpreted in
terms of field-theoretic renormalization \cite{Goldenfeld}.

The same result can in principle be recovered within the present framework
from the RG flow equation, which leads to a nonlinear eigenvalue problem.
This problem was originally derived in a quite different manner and solved 
\textit{numerically} by Barenblatt \cite{Barenblatt}. This example therefore
demonstrates that anomalous scaling, previously treated within
field-theoretic RG, is naturally incorporated into the present unified
framework. Stability analysis based on the RG flow equation has also been
discussed by the present author \cite{okumura2026combined}, while a complete
but much more involved discussion is given in Ch 8.3.2 of \cite{Barenblatt}.

We predict that the \textit{Barenblatt universality class} consists of PDEs
of the form $\partial _{T}H(T,X)=D\partial _{X}^{2}H(T,X)+G$ where the
regular analytic term $G$ is given by a linear combination of terms that can
be expressed as $H^{n_{0}}(\partial _{X}H)^{n_{1}}(\partial
_{X}^{2}H)^{n_{2}}(\partial _{X}^{3}H)^{n_{3}}\cdots $ with a \textit{%
negative }$M=3-(n_{0}+2n_{1}+3n_{2}+\cdots )-2\alpha
(n_{0}+n_{1}+n_{2}+\cdots )$. As an example, a simple nonlinear PDE where $G$
is a linear combination of $H^{n}$ with $n(1+2\alpha )>3$ ($0<\alpha \ll 1$)
shares the same self-similar solution with $\partial _{T}H(T,X)=D\partial
_{X}^{2}H(T,X)$, illustrating the broad scope of the universality class.

\subsubsection{Example 4: Oil-drop PDE}

As already mentioned, the breakup of a moderately viscous liquid in air is
described by Eq \ref{G3D}, which has been well established theoretically 
\cite{1993PRLEggersPinchoff,brenner1994iterated,brenner1996pinching} and
experimentally \cite{1994ScienceNagelDropFallingFaucet}. In this case ($L<1$
and $n=2$), we calculate the number $M_{I}$ for nonzero terms in $\mathbf{F}(%
\mathbf{H},D_{1}\mathbf{H},D_{2}\mathbf{H},\cdots )$ for Eq \ref{G3D}, as in
Appendix \ref{A2b}. The scale-invariance of Eq \ref{c3}, i.e., Eq \ref{G3D},
requires all $M_{I}$ explicitly given in Appendix \ref{A2b} to vanish, which
determines \textit{all the exponents} as%
\begin{equation}
A_{1}=-2\text{, }A_{2}=1\text{, and }B=2\text{,}
\end{equation}%
making this a case of the first kind in Barenblatt's classification. From
the RG FPs in Eq \ref{e20}, we obtain the self-similar solutions: 
\begin{eqnarray}
H^{\ast }(T,X) &=&T^{-A_{1}/B}h^{\ast }(X/T^{1/B})=Th^{\ast }(X/T^{1/2}) \\
V^{\ast }(T,X) &=&T^{-1/2}v^{\ast }(X/T^{1/2})
\end{eqnarray}%
Here, we remark a new feature, $V^{\ast }\rightarrow \infty $ in the limit $%
T\rightarrow 0$ in Case I, while, in the all examples above, $H^{\ast
}\rightarrow 0$ in the limit $T\rightarrow 0$ for Case I and $T\rightarrow
\infty $ for Case II.

From Eq \ref{e11}, by introducing functions $\phi (\tau ,X)\equiv H_{L}(1,X)=%
\emph{R}_{L}f_{1}(X)$ and $\psi (\tau ,X)\equiv V_{L}(1,X)=\emph{R}%
_{L}f_{2}(X)$, we have the RG flow equations: 
\begin{equation}
\left\{ 
\begin{array}{c}
-\frac{d\phi }{d\tau }=-\phi +\frac{\xi }{2}\frac{\partial \phi }{\partial
\xi }+\left( \psi \frac{\partial \phi }{\partial \xi }+\frac{\phi }{2}\frac{%
\partial \psi }{\partial \xi }\right) \text{ } \\ 
-\frac{d\psi }{d\tau }=\frac{\psi }{2}+\frac{\xi }{2}\frac{\partial \psi }{%
\partial \xi }+\left( \psi \frac{\partial \psi }{\partial \xi }+\frac{%
\partial }{\partial \xi }\left( \frac{1}{\phi }\right) -3\frac{1}{\phi ^{2}}%
\frac{\partial }{\partial \xi }\left( \frac{\partial \psi }{\partial \xi }%
\phi ^{2}\right) \right)%
\end{array}%
\right.  \label{FP0}
\end{equation}%
for $\phi =\phi (\tau ,\xi )$ and $\psi =\psi (\tau ,\xi )$.

In the DSD \cite{giga1985asymptotically,eggers2015singularities}, the same
equation is derived in a significantly different manner by introducing (1)
'the logarithmic time'\ $\tau $ by $\tau =-\log T$ and (2) 'the scaling
functions'\ $\phi (\tau ,\xi )$ and $\psi (\tau ,\xi )$ by $H^{\ast }(T,X)$ $%
=$ $T^{\alpha _{1}\text{ }}\phi (\xi =X/T^{\beta },\tau )$ and $V^{\ast
}(T,X)$ $=$ $T^{\alpha _{2}\text{ }}\psi (\xi =X/T^{\beta },\tau )$, without
defining an RG transformation. By contrast, within the present unified RG
framework, Eq~\ref{FP0} emerges naturally from the RG transformation itself,
thereby identifying the DSD flow with an RG flow.

The FPs are obtained by setting the left-hand sides of Eq \ref{FP0} to zero.
This is precisely the analysis carried out by Eggers and others, although
the equation was originally derived in a very different manner. The
resulting studies provided a deep understanding of the singular dynamics and
stability of the oil-drop problem (see Refs. \cite%
{1993PRLEggersPinchoff,1994ScienceNagelDropFallingFaucet,brenner1994iterated,brenner1996pinching}
and Ch.~7.3 of the text \cite{eggers2015singularities}).

We predict that the \textit{Oil-drop breakup universality class} consists of
the \textit{nonlinear systems} 
\begin{equation}
\left\{ 
\begin{array}{c}
\frac{\partial H}{\partial T}=V\frac{\partial H}{\partial X}+\frac{H}{2}%
\frac{\partial V}{\partial X}+G_{H} \\ 
\frac{\partial V}{\partial T}=V\frac{\partial V}{\partial X}+\frac{\partial 
}{\partial X}\left( \frac{1}{H}\right) -3\frac{1}{H^{2}}\frac{\partial }{%
\partial X}\left( \frac{\partial V}{\partial X}H^{2}\right) +G_{V}%
\end{array}%
\right.  \label{ODG}
\end{equation}%
where regular analytic functions $G_{H}$ and $G_{V}$ are given by linear
combinations of terms that can be expressed as $H^{n_{0}}$ $(\partial
_{X}H)^{n_{1}}$ $(\partial _{X}^{2}H)^{n_{2}}$ $(\partial _{X}^{3}H)^{n_{3}}$
$V^{m_{0}}$ $(\partial _{X}V)^{m_{1}}$ $(\partial _{X}^{2}V)^{m_{2}}$ $%
(\partial _{X}^{3}V)^{m_{3}}$ with a \textit{positive }$M_{1}$ and $M_{2}$
defined as $M_{1}=$ $2n_{0}+n_{1}-$ $(n_{3}+2n_{4}+3n_{5}+\cdots )$ $%
-(m_{0}+2m_{1}+3m_{2}+\cdots )$ and $M_{2}$ $=3+2n_{0}+n_{1}$ $%
-(n_{3}+2n_{4}+3n_{5}+\cdots )$ $-(m_{0}+2m_{1}+3m_{2}+\cdots )$. As an
example, a simple nonlinear system in Eq \ref{ODG}, for which $G_{H}$ and $%
G_{V}$ are respectively given by a linear combination of $H^{n}$ and $H^{m}$
with $n>0$ and $2m>-3$ shares the same self-similar solutions with the
oil-drop PDEs in Eq \ref{G3D}, illustrating the broad scope of the
universality class.

\section{Discussion}

Taken together, Examples 1--4 illustrate several complementary aspects of
the framework. In terms of the scale factor, Examples 1 and 4 belong to Case
I ($L<1$), where asymptotic dynamics are governed by \textit{small-scale}
physics. In contrast, Examples 2 and 3 belong to Case II ($L>1$), where 
\textit{large-scale} behavior dominates, as in critical phenomena.

In Barenblatt's classification, Examples 2 and 4 represent self-similarity
of the first kind, for which the similarity exponents follow from
dimensional constraints, including the conservation law in the diffusion
problem, whereas Examples 1 and 3 belong to the second kind and require
additional selection arguments. Interestingly, this distinction parallels
the memory properties of these examples: classical diffusion (Example 2) and
the oil-drop problem (Example 4) lose memory of the initial length scale,
whereas the bubble problem (Example 1) and the Barenblatt problem (Example
3) retain memory associated with an initial length scale. This
correspondence between similarity class and memory retention is discussed
further in Appendix \ref{A-hi}.

Examples 1, 2, and 4 exemplify the unification of BKL-RG and the DSD
achieved by the present framework, whereas Example 3 further extends the
framework to anomalous scaling and thereby establishes its connection to
field-theoretic RG.

Importantly, the linear diffusion FP appearing in Example 2 is shared by a
broad class of \textit{nonlinear} PDEs belonging to the same universality
class. This illustrates the central idea of the present framework: scale
invariance generates self-similar dynamics as RG FPs, whereas repeated RG
transformations eliminate non-scale-invariant deviations. As discussed in
Example 2, the same FP structure also encompasses the stochastic EW equation
and the biological DD diffusion model, providing a common RG description of
deterministic, stochastic, and biological systems.

Previous studies typically start with a given PDE and seek its asymptotic
self-similar solution. The present framework instead asks a complementary
question: given a self-similar fixed point, which PDEs flow toward it under
RG evolution? Once a fixed point is identified, the framework identifies the
corresponding universality class consisting of infinitely many PDEs sharing
the same asymptotic self-similar solution.

Except for the diffusion class, which is partially supported by the rigorous
mathematical results of Bricmont, Kupiainen, and Lin \cite%
{bricmont1994renormalization}, the universality classes identified here do
not appear to have been established previously. They therefore represent
direct consequences of the framework and provide concrete predictions that
can be examined by future numerical simulations and mathematical analyses.

\section{Conclusion \label{Conclusion}}

We developed a unified RG framework for a generic PDE in Eq \ref{e6},
consisting of a scale-invariant $\mathbf{F}$ term and a non-scale-invariant $%
\mathbf{G}$ term, by integrating DSD stability analysis with the BKL-RG and
field-theoretic RG theories. Within this framework, the scale invariance of $%
\mathbf{F}$ generates self-similar solutions as RG FPs, while universality
classes arise through the elimination of irrelevant deviations $\mathbf{G}$
characterized by Eq \ref{e1} under repeated RG transformations. From this
perspective, universality in nonlinear PDEs originates from the same
fundamental RG mechanism that underlies critical phenomena: diverse systems
flow toward a common scale-invariant FP structure under repeated RG
transformations. Such universality helps explain the ubiquity of universal
behavior in nature: although governing equations used in practice are often
approximate, a broad class of deviations becomes irrelevant as the system
approaches the asymptotic regime of interest.

While the present work focuses on the construction of universality classes,
the RG structure of nonlinear PDEs includes several additional aspects that
deserve separate investigation. In particular, anomalous scaling can be
accompanied by retention of initial-condition information, as illustrated by
the Barenblatt problem above and further discussed from the viewpoint of RG
fixed-point structure in Ref \cite{Okumura2026memory} (we return to this
issue later). Traveling waves and velocity selection had already been
treated using an iterative Wilson-type RG by Chen and Goldenfeld \cite%
{chen1995numerical}. Ref \cite{Okumura2026traveling} addresses a more
specific question, namely whether such traveling-wave fixed points generate
universality classes through the usual RG mechanism of eliminating
irrelevant structures. Ref \cite{Okumura2026FKPP} similarly examines the
distinct question of whether RG fixed-point stability determines front
selection.

The framework unifies finite-time singularities, long-time intermediate
asymptotics, stochastic growth processes, and biological diffusion systems
within a common RG description, clarifying the distinct roles of small-scale
physics for $L<1$ and\ large-scale physics for $L>1$. In doing so, it
provides a constructive procedure for identifying scale invariance, FPs,
stability, and universality class, on the basis of the scaling exponent
defined in Eq \ref{e1} and the RG flow equation in Eq \ref{e11}. The
predictive nature of the framework is illustrated by the universality
classes that emerge directly from the RG analysis.

As in critical phenomena, scaling behavior in PDE dynamics should be
understood in relation to the asymptotic RG regime being probed. A
self-similar fixed point may govern either an intermediate scaling regime or
the ultimate asymptotic behavior. In the former case, a term that is
negligible over an intermediate range may eventually become important and
induce crossover toward another fixed point or asymptotic state. A
dynamically stable fixed point does not, however, imply convergence from
arbitrary initial conditions. Convergence depends on whether the RG
trajectory lies within the basin of attraction of the corresponding fixed
point. Nevertheless, in many fluid-interface singularities, dynamically
stable similarity solutions are approached from a range of initial
conditions without additional tuning of external parameters \cite%
{1993PRLEggersPinchoff,1994ScienceNagelDropFallingFaucet,brenner1994iterated,brenner1996pinching}%
. Thus, dynamical stability remains important for understanding which
self-similar states are physically realized, but such attraction does not
imply that a given fixed point must be the ultimate global attractor. The
relation among stability, observability, and crossover in PDE dynamics is
discussed further in Appendix \ref{A4}.

Although PDE-based systems are mathematically more tractable than critical
phenomena, \textit{stability of self-similar solutions themselves} remains
technically demanding even in relatively simple settings. Accordingly, we
restrict explicit stability analysis here to the simplest case of linear
diffusion, which nevertheless remains nontrivial (see Appendix \ref{A3}). As
in critical phenomena, however, scaling exponents and the associated RG flow
behavior are often sufficient for the interpretation of experiments. The
present framework therefore provides a practical methodology through the
systematic derivation of scale-invariance conditions, RG flow equations, and
universality classes, particularly for self-similarity of the first kind,
while remaining broadly applicable to self-similarity of the second kind.

As discussed above, anomalous scaling is often accompanied by retention of
initial-condition information. An important next step is to extend the
present predictive framework to such systems, building on previous
developments \cite{Goldenfeld, goldenfeld1989intermediate,
goldenfeld1990anomalous, chen1994renormalization, chen1995numerical}. In
such systems, a central question is which initial or system scales survive
in the asymptotic fixed point. In the Barenblatt problem, for example, the
initial width $l$, which is lost in classical diffusion, remains encoded
through $m_{0}l^{2\alpha }$. Ref \cite{Okumura2026memory} extends this
viewpoint by classifying RG fixed points according to how such information
survives asymptotically. A remaining question is how selective memory
retention can be incorporated into the present predictive construction of
universality classes.

The experimental observations discussed in Appendix \ref{A-hi} make this
question particularly relevant when multiple system scales are present.
While the RG theory developed here classifies PDEs according to the
asymptotic self-similar solutions they share, these experiments reveal an
additional hierarchy associated with the extent to which asymptotic dynamics
retain information about different system scales. Understanding how memory
loss, memory retention, and partial memory emerge from scale separation and
RG flow therefore remains an important direction for future work.

A related challenge is to uncover new self-similar solutions of the
Navier--Stokes equations. Our recent experiments in confined fluid systems 
\cite{YokotaPNAS2011,nakazato2018self,nakazato2022air,yoshino2025partial}
have shown that controlled variations of external conditions, such as
geometric confinement, can lead to distinct self-similar dynamics,
suggesting the existence of multiple universality classes. These findings
point to a rich landscape of asymptotic behaviors whose systematic
organization remains an open problem and whose governing equations are often
unknown. The present RG framework may help identify such solutions in a
systematic manner.

More broadly, the generic form in Eq \ref{e1} encompasses a wide class of
PDE-driven systems beyond the examples considered here, including chemical
reactions, biological rhythms, and nonequilibrium pattern formation \cite%
{murray2002mathematical1, cross1993pattern}. Given the widespread occurrence
of self-similar dynamics across natural phenomena and industrial processes,
from geological and magmatic systems to petroleum engineering and
microfluidic technologies \cite%
{parmigiani2016bubble,StoneStroockAjdari2004,HeleShawPetroleum2010,anna2016}%
, the unified RG framework developed here provides a systematic foundation
for understanding and classifying universality in a broad range of nonlinear
dynamical systems.

Finally, we emphasize that the universality classes constructed in the
present work follow directly from the RG analysis developed here and provide
a set of concrete predictions that can be tested by future numerical
simulations and mathematical analyses.

\textit{The author is grateful to Professor Nigel Goldenfeld (UCSD) for
bringing several important references to the author's attention and for
providing warm encouragement. This work was supported by JSPS\ KAKENHI Grant
Number JP24K00596.}

Data Availability Statements: Data sharing is not applicable to this article
as no new data were created or analyzed in this study.

\appendix

\section*{Appendix: Supplementary Discussions}

\subsection{Hierarchical structure of universality\label{A-hi}}

\begin{figure}[tbp]
\includegraphics[width=0.9\columnwidth]{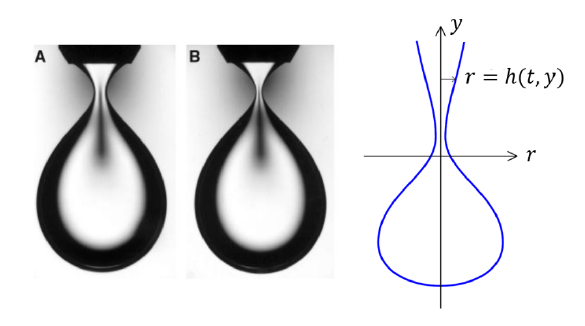}
\caption{Left: Formation of a 'bubble' (reproduced from Doshi and Nagel et
al., Science 302 (2003) with permission from AAAS). Right: Shape function $%
h(t,y)$.}
\label{Fig1}
\end{figure}


Experimental observations suggest a useful hierarchy in the manifestations
of universality, which we illustrate using fluid-interface topological
transitions (Fig 2 \cite{2003ScienceNagelMemoryDropBreakup}) \cite%
{1994ScienceNagelDropFallingFaucet,hernandez2012symmetric,YokotaPNAS2011,nakazato2018self,nakazato2022air,yoshino2025partial,pahlavan2019restoring,kaneelil2022three}%
. This experimentally motivated hierarchy is not intended as an alternative
classification of self-similar solutions to the classical classification
introduced by Barenblatt \cite{barenblatt2003scaling, barenblatt2003scaling}%
. Rather, it organizes experimental observations according to the amount of
scaling information shared across different times and experimental
conditions.

As seen in Figs 3 and 4 \cite{YokotaPNAS2011,nakazato2018self},
appropriately rescaled coalescence and breakup dynamics collapse onto 
\textit{master curves} $Y=\Gamma (X)$, reflecting the emergence of \textit{%
characteristic scales} $r(t)$ (or, more generally, different scales $T^{a}$
and $T^{b}$ in Eq A of Fig 3). The existence of such \textit{data collapse},
irrespective of the specific form of $\Gamma (X)$, defines the \textit{%
zeroth level }of universality, which is observed across diverse nonlinear
systems in nature. At a higher level, for a given phenomenon under fixed
experimental conditions the master curve $Y=\Gamma (X)$ is unique and
defines a specific asymptotic self-similar solution. This defines the 
\textit{first level} of universality, reflecting the universality for change
in time. At a still higher level, a stronger form of universality arises
when the same master curve persists under variations of experimental
conditions (see the bottom row of Fig 4).

This \textit{second level} of universality is closely related to whether the
dynamics \textit{retain memory} of system parameters. When several
independent system scales are present, the second level can itself exhibit
different degrees, depending on how many of these scale dependences
disappear asymptotically. For a system characterized by three independent
scales, for example, three degrees can be distinguished within the second
level: dependence may remain on two scales, on only one scale, or on none of
them, with the last case corresponding to complete second-level
universality. Our confined-interface experiments provide an example of the
intermediate case: among three distinct geometrical length scales, the
asymptotic master curve retains dependence on the smallest scale while
becoming independent of the other two \cite{nakazato2018self,
yoshino2025partial}. We refer to this selective retention of only part of
the system-scale information as \textquotedblright partial
memory.\textquotedblright\ 

The familiar binary distinction between memory loss and memory retention can
then be viewed as the special case in which only one length scale carries
the relevant memory. In this simpler setting, the relation to Barenblatt's
classification becomes particularly transparent. Classical diffusion and the
Barenblatt problem, discussed in Examples 2 and 3, respectively, provide
canonical examples. In classical diffusion, a self-similar solution of the
first kind, the conserved mass $m_{0}${} remains in the long-time asymptotic
solution, whereas the initial width $l$ disappears. In the Barenblatt
problem, by contrast, self-similarity is of the second kind and information
about $l$ survives in the anomalous amplitude through the combination $%
m_{0}l^{2\alpha }$, as shown explicitly in Eq (\ref{H0}). Thus, these two
diffusion problems reveal a close correspondence between first- versus
second-kind similarity and the loss versus retention of information about a
single initial length scale \cite{Barenblatt, Goldenfeld}.

The same correspondence is suggestive in other examples considered here. The
oil-drop problem (Example 4) exhibits universal self-similar behavior and
belongs to the first kind, whereas the bubble problem (Example 1), which
exhibits persistence of memory, belongs to the second kind, as summarized in
Sec. III. The converging gravity current discussed in Sec. II.C.3 provides
another example of the second kind in which a dimensionless parameter
selected by the initial condition remains encoded in the asymptotic solution 
\cite{diez1992self, verga2003elements}. Taken together, these examples
suggest that the familiar first-/second-kind distinction is closely related
to a binary loss-or-retention distinction when the asymptotic solution
retains memory of a single parameter associated with the initial condition.
This connection has been explored further from an RG fixed-point perspective
in Ref. \cite{Okumura2026memory}.

The multiple-scale experiments therefore raise a question that goes beyond
this simple binary setting. When several system scales are present, some
scale information can be asymptotically lost while other information is
retained, giving rise to intermediate degrees of universality. Whether such
selective memory retention can be classified in terms of self-similarity of
the first or second kind remains unknown, because the governing PDEs for
these experimentally observed self-similar regimes have not yet been
established. Determining whether the experimental hierarchy can ultimately
be related to Barenblatt's classification through a more general relation
among similarity class, scale separation, and memory retention remains an
open problem.

\begin{figure*}[tbp]
\includegraphics[width=0.9\textwidth]{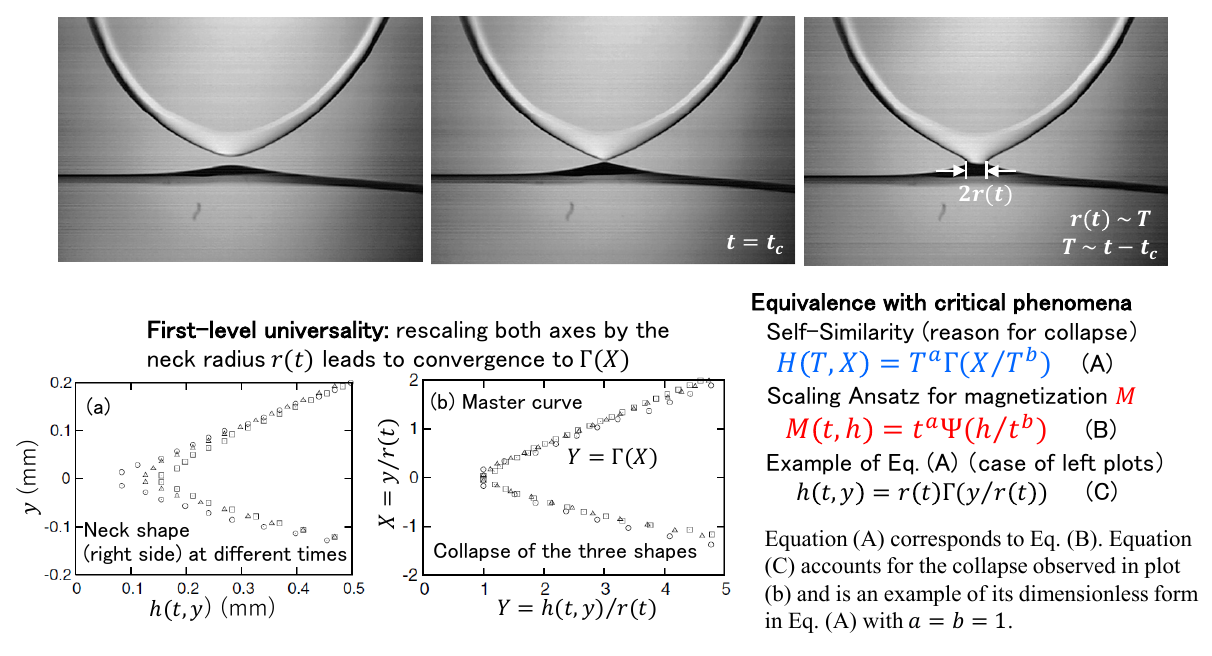}
\caption{Coalescence of a liquid drop in a confined liquid-liquid system
under an electric field. [Top image] Under the influence of the electric
field, the lower tip of the droplet sharpens and coalesces with the
underlying bulk phase. [Bottom graphs] (a): Interfacial profile near the
right-side neck region. (b): Convergence of the three interfacial shapes
from (a) on rescaled axes (data collapse), demonstrating the first-level
self-similarity. Images and graphs are adapted from Maria Yokota and Ko
Okumura, Proc. Natl. Acad. Sci. U.S.A. 108 (2011). Copyright 2011 M. Yokota
and K. Okumura.}
\label{Fig2}
\end{figure*}

\subsection{Note for bridging the gap between fields\label{A0}}

\textit{Self-similar solution and scaling ansatz-- }Self-similar solutions
play a central role in the description of singular dynamics and intermediate
asymptotics in a wide range of systems governed by partial differential
equations (PDEs). These solutions are typically expressed in the form given
in Eq \ref{SS}, which implies the existence of characteristic scales
governed by power laws. Such scaling forms closely parallel the well-known
scaling ansatz in critical phenomena in Eq \ref{SA}, where physical
quantities near critical points exhibit universal behavior characterized by
scaling functions and critical exponents. This analogy has long suggested
that similar underlying principles may connect self-similar dynamics in PDEs
with universality in statistical physics.

\textit{Perspective from the dynamical-system description-- }Within the
framework of the dynamical-system description (DSD), self-similar solutions
have been introduced and analyzed with remarkable success. In this approach,
the existence of self-similarity is typically taken as a working assumption,
which enables the reduction of complex PDE dynamics to tractable dynamical
systems, allowing stability analysis and classification of solutions. This
methodology has led to substantial progress in understanding singular
dynamics, particularly in fluid systems and related areas, and has provided
a powerful and widely accepted framework for analyzing experimental
observations.

\textit{Complementary role of the renormalization-group perspective-- }The
present work builds on this established DSD by introducing a complementary
perspective based on renormalization-group (RG) theory. In contrast to
approaches that take self-similar forms as a starting point for analysis,
the RG framework provides a systematic procedure in which such structures
emerge naturally as fixed points (FPs) of the RG transformation. In this
formulation, self-similar solutions are understood as a direct consequence
of scale invariance of the governing equations. This perspective is closely
aligned with the role played by RG theory in critical phenomena, where the
scaling ansatz---historically introduced phenomenologically---was later
understood as arising from the structure of RG FPs.

\textit{Bridging two conceptual frameworks-- }In other words, the DSD and RG
approaches can be seen as complementary rather than competing descriptions:
The DSD framework provides an effective and practical methodology for
identifying and analyzing self-similar solutions and their stability in
specific systems. The RG framework offers a unifying theoretical foundation
that explains why such self-similar structures arise and how they are
organized into universality classes. By combining these perspectives, the
present work aims to bridge two traditions that have developed largely
independently, and to provide a common conceptual framework for
understanding universality in dynamical systems.

\textit{Relevance of the present work--} Because the conceptual contribution
of this work lies in clarifying the origin of self-similar
structures---rather than in identifying new solutions within a specific
model---it may differ in emphasis from studies that focus primarily on
detailed dynamical analysis within the DSD framework. We therefore hope that
this supplementary note may assist in placing the present results in a
broader conceptual context, particularly for readers whose expertise is
rooted in one of the two traditions described above.

\begin{figure*}[tbp]
\includegraphics[width=\textwidth]{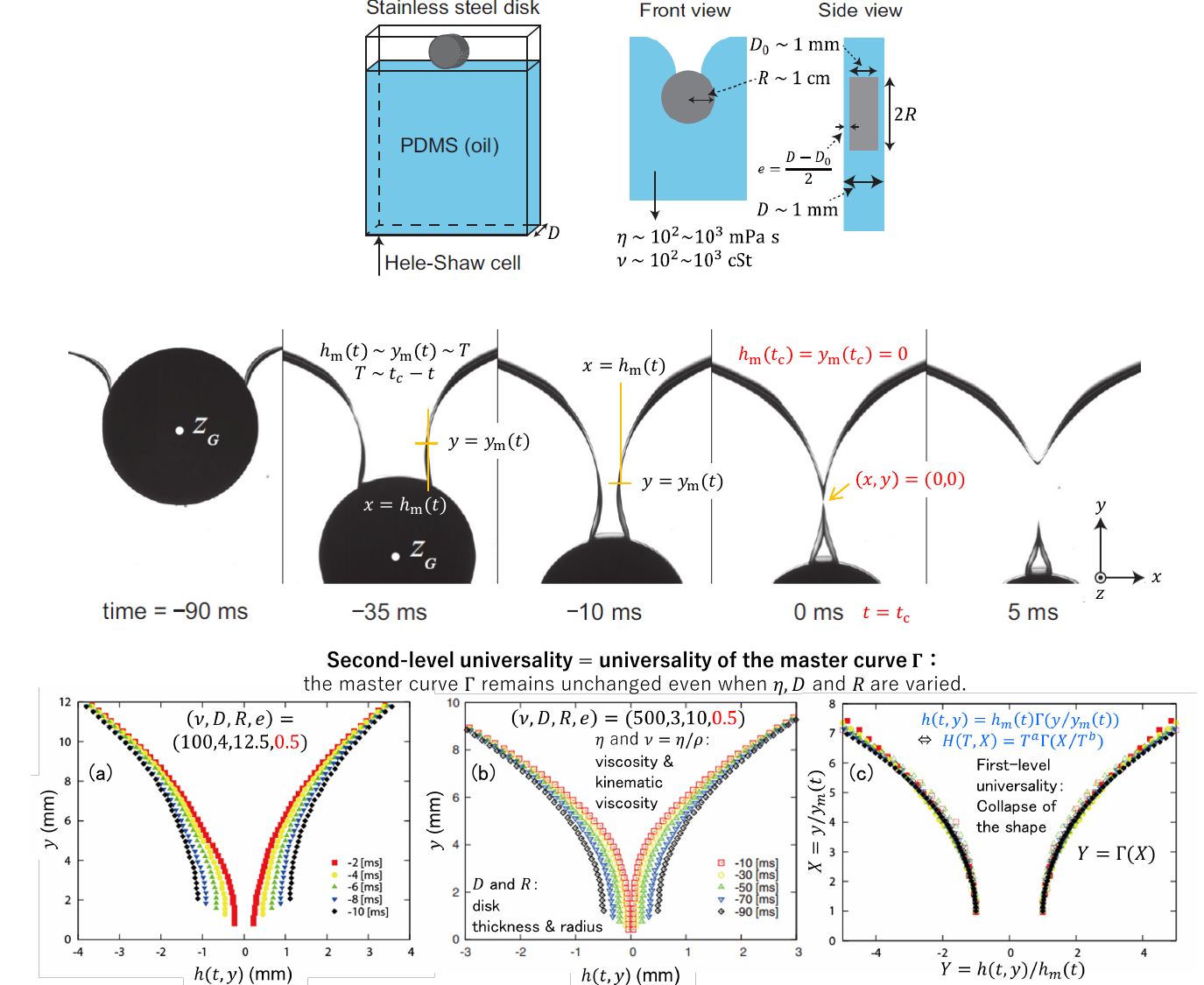}
\caption{Breakup of a bubble in a confined geometry. [Top illustrations]
Left: Experimental setup. Right: The front view shows a disk entraining air
into the liquid phase, while the side view illustrates the formation of thin
films of thickness $e$ between the disk and the wall of the cell. [Middle
snapshots] A frontal view showing the disk entraining air into the liquid
phase and eventually causing it to pinch off. [Bottom graphs] (a): Time
evolution of the interfacial shape above the point of minimum neck radius.
(b): Interfacial profiles under different experimental conditions. (c): The
ten profiles from (a) and (b) plotted on rescaled axes, demonstrating
collapse onto a single master curve. Adapted from Hana Nakazato, Yuki
Yamagishi, and Ko Okumura, Phys. Rev. Fluids 3 (2018). (CC BY 4.0).
Copyright 2018 H. Nakazato, Y. Yamagishi, and K. Okumura. }
\label{Fig3}
\end{figure*}

\subsection{Mathematical Details\label{A2}}

\subsubsection{Full expression for $\mathbf{F}$\label{A2a}}

The $I$-th component of $\mathbf{F}$ can be written as $F_{I}(\mathbf{H}%
,D_{1}\mathbf{H},D_{2}\mathbf{H},\cdots )$ $=$ $\sum_{\{N_{i}\},\{N_{jk}\},%
\{N_{lmn}\}\cdots }$ $f_{\{N_{i}\},\{N_{jk}\},\{N_{lmn}\}\cdots \text{ }%
}^{I}\prod\limits_{i,j,k,l,m,n,\cdots }$ $H_{i}^{N_{i}}$ $(\partial
_{X_{j}}H_{k})^{N_{jk}}$ $(\partial _{X_{l}}\partial
_{X_{m}}H_{n})^{N_{lmn}}\cdots $, where only a few coefficients $%
f_{\{N_{i}\},\{N_{jk}\},\{N_{lmn}\}\cdots }^{I}$ are nonzero. For $n=2$ and $%
m=2$, where $\mathbf{H}=(H_{1},H_{2})$ and $\mathbf{X}=(X_{1},X_{2})$, the
coefficient $f^{I}$ is given as $f_{\{N_{i}\},\{N_{jk}\},\{N_{lmn}\}\cdots
}^{I}$ $=$ $%
f_{N_{1}N_{2},N_{11}N_{12}N_{21}N_{22},N_{111}N_{112}N_{121}N_{122}N_{211}N_{212}N_{221}N_{222}}^{I}. 
$

\subsubsection{Derivation of the general expression of $M_{I}$ in Eq. (%
\protect\ref{e1})\label{A-D1}}

By making the replacements, $\mathbf{X}\rightarrow L\mathbf{X}$ and $%
T\rightarrow L^{B}T$, in the left-hand side of Eq. (\ref{c3}) and then using
Eq. (\ref{Ar3c}), the term $\frac{\partial H_{i}(L^{B}T;L\mathbf{X})}{%
\partial L^{B}T}$ is expressed as $\frac{\partial H_{i}^{L}(T;\mathbf{X}%
)/L^{A_{i}}}{L^{B}\partial T}$ $=$ $L^{-A_{i}-B}$ $\frac{\partial
H_{i}^{L}(T;\mathbf{X})}{\partial T}$. Similar consideration for the term $%
H_{i}^{N_{1}}(\partial _{X_{j}}H_{k})^{N_{jk}}(\partial _{X_{l}}\partial
_{X_{m}}H_{n})^{N_{lmn}}\cdots $, we obtain $\left[ H_{i}(L^{B}T;L\mathbf{X})%
\right] ^{N_{1}}$ $\left[ \frac{\partial H_{k}(L^{B}T;L\mathbf{X})}{\partial
LX_{j}}\right] ^{N_{jk}}$ $\left[ \frac{\partial ^{2}H_{n}(L^{B}T;L\mathbf{X}%
)}{\partial LX_{l}\partial LX_{m}}\right] ^{N_{lmn}}\cdots
=L^{-N_{i}A_{i}-N_{ij}(A_{k}+1)-N_{lmn}(A_{n}+2)}$ $\left[ H_{i}^{L}(T;%
\mathbf{X})\right] ^{N_{1}}$ $\left[ \frac{\partial H_{k}^{L}(T;\mathbf{X})}{%
\partial X_{j}}\right] ^{N_{jk}}$ $\left[ \frac{\partial ^{2}H_{n}^{L}(T;%
\mathbf{X})}{\partial X_{l}\partial X_{m}}\right] ^{N_{lmn}}\cdots $. From
these relations, we can understand Eq \ref{e1}.

\subsubsection{Calculation of $M_{I}$\label{A2b}}

In Example 1, $\mathbf{F}(\mathbf{H}_{L},D_{1}\mathbf{H}_{L},D_{2}\mathbf{H}%
_{L},\cdots )\rightarrow 1$ and the only nonzero coefficient $%
f^{I}\rightarrow f_{N_{1}N_{11}N_{111}\cdots }=f_{00\cdots \cdots }=1$. This
leads to $M=A+B$. More generally, the term in Eq \ref{t1} in this case is
given by $H^{n_{0}}(\partial _{X}H)^{n_{1}}(\partial
_{X}^{2}H)^{n_{2}}(\partial _{X}^{3}H)^{n_{3}}$ so that $M$ $=$ $A+B-$ $%
[n_{0}A+n_{1}(A+1)+n_{2}(A+2)+\cdots ]$, i.e., 
\begin{equation}
M=A+B-A(n_{0}+n_{1}+n_{2}+\cdots )-(n_{1}+2n_{2}+3n_{3}+\cdots ),  \label{M1}
\end{equation}%
from which we understand the expression of $M$ for the extra $G$ term by
noting $A=-B=-2$.

In Example 2, $\mathbf{F}(\mathbf{H}_{L},D_{1}\mathbf{H}_{L},D_{2}\mathbf{H}%
_{L},\cdots )\rightarrow \partial _{X}^{2}H(T,X)$ and the only nonzero
coefficient $f^{I}\rightarrow f_{N_{1}N_{11}N_{111}\cdots }=f_{00100\cdots
}=1$. This leads to $M=A+B-(A+2)$. The term in Eq \ref{t1} in this case is
given as in Example 1, we obtain $M=3-(n_{0}+2n_{1}+3n_{2}+\cdots )$ from Eq %
\ref{M1} by noting $A=1$ and $B=2$. In Example 3, we instead obtain $M$ with
the extra $\alpha $ term as in the text.

In the case of the noisy EW equation, we introduce the scale transformation
for the stochastic variable as 
\begin{eqnarray}
\Theta ^{\prime }(T^{\prime },X^{\prime }) &=&L^{\alpha }\Theta (T,X)\
\equiv \Theta ^{L}(T^{\prime },X^{\prime })  \notag \\
&\Leftrightarrow &\Theta ^{L}(T,X)=L^{\alpha }\Theta (L^{B}T,LX).
\label{e30}
\end{eqnarray}%
We determine the exponent $\alpha $ from the scale invariance of the noise
correlation, $\left\langle \Theta (T_{1},X_{1})\Theta
(T_{2},X_{2})\right\rangle $ $=\delta (T_{1}-T_{2})\delta (X_{1}-X_{2})$. By
making the transformation, $T\rightarrow L^{B}T$ and $X\rightarrow LX$, we
obtain $\left\langle \Theta (L^{B}T,LX)\text{ }\Theta (0,0)\right\rangle $ $%
=\delta (L^{B}T)\delta (LX)$, which leads to $L^{-2\alpha }\left\langle
\Theta ^{L}(T,X)\Theta ^{L}(0,0)\right\rangle $ $=L^{-B-1}\delta (T)\delta
(X)$ by using Eq \ref{e30} and $\delta (cx)=\delta (x)/c$. Thus, the scale
invariance require $2\alpha =B+1$. For the spatial diffusion term $\partial
_{X}^{2}H$ (with $N_{111}=1$), the dimension $M_{1}=A+B-(A+2)=B-2$, while
for the noise term $\Theta (T,X)$, it is given by $M_{1}=A+B-(B+1)/2$. To
ensure the simultaneous scale invariance of both terms, we require $B=2$ and 
$A=-1/2$, and thus we obtain the EW scaling, $H^{\ast }(T,X)$ $=$ $%
T^{1/4}h_{i}^{\ast }(X/T^{1/2})$, from Eq \ref{e20}. We predict that the 
\textit{EW universality class} consists of the noisy EW equation with the
regular analytic term $G$ given by a linear combination of terms that can be
expressed as $H^{n_{0}}(\partial _{X}H)^{n_{1}}(\partial
_{X}^{2}H)^{n_{2}}(\partial _{X}^{3}H)^{n_{3}}\cdots $ with a \textit{%
negative} $M$ in Eq \ref{M1} with $B=2$ and $A=-1/2$. For example, the term $%
\left( \partial _{X}^{2}H\right) ^{2}$ is an irrelevant term, while the
nonlinear KPZ term, $(\partial _{X}H)^{2}$, is relevant.

In the case of the biological DD diffusion model, for the two terms from the
diffusion term $\partial _{X}(H^{m}\partial _{X}H)$, with $%
(N_{1},N_{11})=(m-1,2)$ and $(N_{1},N_{111})=(m,1)$, the dimensions are
respectively given as $M_{1}=A+B-(m-1)A-2(A+1)$ and $A+B-mA-(A+2)$, both
reducing to the same value $-mA+B-2$. To ensure the scale invariance of both
terms, we require $B=mA+2$. If we further require the invariance of the
conservation law $\int dXH=1$, for which $\int dXH\rightarrow $ $\int
dLXH(LT^{B},LX)=L^{1-A}\int dXH$, to obtain $A=1$, i.e., $B=2+m$. Thus we
obtain the DD diffusion scaling, $H^{\ast }(T,X)$ $=$ $T^{-1/(2+m)}h_{i}^{%
\ast }(X/T^{1/(m+2)})$, from Eq \ref{e20}. We predict that this\textit{\
universality class} consists of the equation with the regular analytic term $%
G$ given by a linear combination of terms that can be expressed as $%
H^{n_{0}}(\partial _{X}H)^{n_{1}}(\partial _{X}^{2}H)^{n_{2}}(\partial
_{X}^{3}H)^{n_{3}}\cdots $ with a \textit{negative} $M$ in Eq \ref{M1} with $%
B=m+2$ and $A=1$.

In Example 4, $M_{I}$ is calculated as follows. For $I=1$, from Eq \ref{G3D}%
, we obtain, for the terms, $\frac{H}{2}\frac{\partial V}{\partial X}$ with $%
N_{1}=N_{12}=1$ and $V\frac{\partial H}{\partial X}$ with $N_{2}=N_{11}=1$, 
\begin{eqnarray}
M_{1} &=&A_{1}+B-(A_{1}+A_{2}+1)=B-A_{2}-1  \label{e21} \\
M_{1} &=&A_{1}+B-(A_{1}+1+A_{2})=B-A_{2}-1
\end{eqnarray}%
As for $I=2$, the second equation in Eq \ref{G3D} can be expressed as $-%
\frac{\partial V}{\partial T}+V\frac{\partial V}{\partial X}=\frac{1}{H^{2}}%
\frac{\partial H}{\partial X}+3\frac{\partial ^{2}V}{\partial X^{2}}+6\frac{%
\partial V}{\partial X}\frac{1}{H}\frac{\partial H}{\partial X}$.
Accordingly, for the four terms, (1) $-\frac{1}{H^{2}}\frac{\partial H}{%
\partial X}$ with $N_{1}=-2$ and $N_{11}=1$, (2) $V\frac{\partial V}{%
\partial X}$ with $N_{2}=N_{12}=1$, (3) $-3\frac{\partial ^{2}V}{\partial
X^{2}}$ with $N_{112}=1$, and (4) $-6\frac{\partial V}{\partial X}\frac{1}{H}%
\frac{\partial H}{\partial X}$ with $N_{1}=-1,N_{11}=1$, and $N_{12}=1$, we
respectively obtain 
\begin{eqnarray}
M_{2} &=&A_{2}+B-(-2A_{1}+A_{1}+1)=B+A_{1}+A_{2}-1  \label{e22} \\
M_{2} &=&A_{2}+B-(A_{2}+A_{2}+1)=B-A_{2}-1 \\
M_{2} &=&A_{2}+B-(A_{2}+2)=B-2 \\
M_{2} &=&A_{2}+B-(-A_{1}+A_{1}+1+A_{2}+2)=B-2
\end{eqnarray}

The term in Eq \ref{t1} in this case is given by $H^{n_{0}}(\partial
_{X}H)^{n_{1}}(\partial _{X}^{2}H)^{n_{2}}(\partial _{X}^{3}H)^{n_{3}}$ $%
V^{m_{0}}$ $(\partial _{X}V)^{m_{1}}$ $(\partial _{X}^{2}V)^{m_{2}}$ $%
(\partial _{X}^{3}V)^{m_{3}}$ so that $M_{I}=$ $A_{I}+B$ $-$ $%
[n_{0}A_{1}+n_{1}(A_{1}+1)+n_{2}(A_{1}+2)+\cdots ]$ $-$ $[m_{0}A_{2}$ $%
+m_{1}(A_{2}+1)$ $+m_{2}(A_{2}+2)+\cdots ]$ with $A_{1}=-2,A_{2}=1$, and $%
B=2 $, from which we obtain the expressions for $M_{1}$ and $M_{2}$ in the
text.

\subsection{Stability analysis for Example 2\label{A3}}

A differential equation obtained by the substitution explained in the text
can be transformed by $\delta $ $=$ $e^{-\xi ^{2}/4}$ $D(\xi )$ into Weber
differential equation for $D(\xi )$ with $\nu =-2\omega $. By requiring a
convergent condition for $\delta (\xi )$ at infinity, we end up with an 
\textit{eigenvalue problem}, obtaining a set of solutions described by
Hermite polynomials: $\delta (\xi )\simeq e^{-\xi ^{2}/2}H_{n}(\xi )$ with $%
-2\omega =n$ with $n=0,1,2,3\cdots $, as in the case of quantum mechanics of
the harmonic oscillator \cite{Schiff1968} (see \cite{okumura2026combined}
for the details). Thus, near the FP, the dynamics is described by a
superposition of them: $\widehat{f}(\xi ,\tau )$ $=$ $f^{\ast }(\xi )$ $%
+\sum_{n=0}^{\infty }$ $a_{n}$ $e^{\omega _{n}\tau }$ $e^{-\xi ^{2}/2}$ $%
H_{n}(\xi )$ with $\omega _{n}$ $=-n$ with $n=0,1,2,\cdots $. If $\omega _{n}
$ are all negative, all the modes decay away by the factor $e^{\omega
_{n}\tau }$ with (logarithmic) "time" because $\tau $ is positive and grows
with "time," and thus the FP solution is stable. In the present case,
although we have a non-negative mode, the $n=0$ mode $\simeq $ $e^{-\xi
^{2}/2}$, called a \textit{marginal mode}, this term can be absorbed into
the FP solution $f^{\ast }(\xi )\simeq $ $e^{-\xi ^{2}/2}$ and thus made to
vanish: We conclude the above self-similar solution is stable. We note a
much more complicated example of the stability analysis for Barenblatt's PDE
is given in Ch 8.3.2 of \cite{Barenblatt}, while a less rigorous version is
given in the Appendix of \cite{okumura2026combined}.

\subsection{Stability, observability, and crossover in PDE dynamics\label{A4}%
}

It is useful to clarify the roles of stability and crossover in the RG
description of PDE dynamics, particularly in relation to the physical
observability of self-similar solutions. A structural difference in the
present RG framework for PDEs and conventional critical phenomena is
particularly important in this context. In the latter, the RG transformation
acts on a Hamiltonian or effective theory, while physical observables such
as the order parameter are obtained from the resulting statistical ensemble.
In the present PDE framework, by contrast, the evolving field $\mathbf{H}(T;%
\mathbf{X})$ is transformed directly under RG and simultaneously represents
the observable dynamical state. In this limited sense, $\mathbf{H}(T;\mathbf{%
X})$ combines aspects that are represented separately by the Hamiltonian and
the order parameter in conventional statistical-mechanical RG.

This structural feature makes dynamical stability especially relevant to
observability in PDE problems. Both stable and unstable self-similar fixed
points may exist, but a dynamically attractive fixed point can be approached
from a range of initial conditions. In many fluid-interface singularities,
experimentally relevant similarity solutions are dynamically robust \cite%
{1993PRLEggersPinchoff,1994ScienceNagelDropFallingFaucet,brenner1994iterated,brenner1996pinching}%
. Thus, stability analysis provides important information about which of the
mathematically admissible self-similar states can be dynamically realized.

Dynamical attraction over a given range of scales, however, does not imply
that the corresponding fixed point is the ultimate global attractor. A term
that is negligible in an intermediate regime may become important at longer
times or different scales, causing crossover toward another fixed point or
asymptotic state. In this respect, PDE dynamics share the general RG
structure of crossover familiar from critical phenomena. Stability and
observability should therefore be understood relative to the scaling regime
under consideration.

\end{document}